\documentclass[aps, prl, reprint, amsmath, amssymb]{revtex4-2}
\usepackage{enumitem}
\usepackage{tabularx}
\usepackage{longtable}
\usepackage{mathrsfs}
\usepackage{lineno}
\usepackage{enumerate}
\usepackage{booktabs}
\usepackage{multirow}
\usepackage{siunitx}
\usepackage[usenames,dvipsnames]{xcolor}
\usepackage{flushend}
\usepackage{graphicx}
\usepackage{natbib}
\usepackage{array}
\usepackage{multirow}
\usepackage[separate-uncertainty=true]{siunitx}
\usepackage{graphicx}
\usepackage{dcolumn}
\usepackage{bm}

\begin{document}
\preprint{APS/123-QED}

\title{Single-event neutron time-of-flight spectroscopy with a petawatt-laser-driven neutron source}

\author{M.~A.~Millán-Callado$^{1,2,\dagger}$}
\author{S.~Scheuren$^{3,\dagger}$}
\author{A.~Alejo$^{4}$}
\author{J.~Benlliure$^{4,5}$}
\author{R.~Beyer$^{6}$}
\author{T.~E.~Cowan$^{6,10}$}
\author{B.~Fernández$^{1,2}$}
\author{E.~Griesmayer$^{7,8}$}
\author{A.~R.~Junghans$^{6}$}
\author{J.~Kohl$^{3}$}
\author{F.~Kroll$^{6}$}
\author{J.~Metzkes-Ng$^{6}$}
\author{I.~Prencipe$^{6}$}
\author{J.~M.~Quesada$^{2}$}
\author{M.~Rehwald$^{6}$}
\author{C.~Rödel$^{3,9}$}
\author{T.~Rodríguez-González$^{1,2}$}
\author{U.~Schramm$^{6,10}$}
\author{M.~Roth$^{3,11}$}
\author{R.~\v{S}tefan\'{\i}kov\'a$^{6,10}$}
\author{S.~Urlass$^{6}$}
\author{C.~Weiss$^{7,8}$}
\author{K.~Zeil$^{6}$}
\author{T.~Ziegler$^{6}$}
\author{C.~Guerrero$^{1,2,*}$}

\affiliation{$^1$Centro Nacional de Aceleradores (CNA), US-Junta de Andalucía-CSIC, 41092 Seville, Spain}
\affiliation{$^2$Dpt. Física Atómica, Molecular y Nuclear (FAMN), Facultad de Física. Universidad de Sevilla (US), 41012 Seville, Spain}
\affiliation{$^3$Technical University of Darmstadt, Institute for Nuclear Physics, 64289 Darmstadt, Germany}
\affiliation{$^4$Instituto Galego de Física de Altas Enerxías (IGFAE), Universidad de Santiago de Compostela, 15782 Santiago de Compostela, Spain}
\affiliation{$^5$Instituto de Física Corpuscular (IFIC), CSIC - Universitat de València, 46980 Valencia, Spain}
\affiliation{$^6$Helmholtz-Zentrum Dresden -- Rossendorf (HZDR), 01328 Dresden, Germany}
\affiliation{$^7$CIVIDEC Instrumentation GmbH, 1010 Vienna, Austria}
\affiliation{$^8$Technische Universität Wien, 1040 Vienna, Austria}
\affiliation{$^{9}$University of Applied Sciences Schmalkalden, 98574 Schmalkalden, Germany}
\affiliation{$^{10}$TUD Dresden University of Technology,  01062, Dresden, Germany}
\affiliation{$^{11}$Focused Energy GmbH,  64293, Darmstadt, Germany}
\affiliation{}
\affiliation{$*$ Corresponding author (cguerrero4@us.es)}
\affiliation{$^{\dagger}$ These authors contributed equally.}

\begin{abstract}
Fast neutron-induced nuclear reactions are crucial for advancing our understanding of fundamental nuclear processes, stellar nucleosynthesis, and applications, including reactor safety, medical isotope production, and materials research. With many research reactors being phased out, compact accelerator-based neutron sources are becoming increasingly important. Laser-driven neutron sources (LDNSs) offer unique advantages---ultrashort neutron pulsees for superior energy resolution, high per-pulse flux, and a drastically reduced footprint. 
However, their use in single-event fast neutron spectroscopy remains unproven, requiring stable multi-shot operation and detectors capable of functioning in the extreme environment of petawatt-class laser-plasma interactions. Here, we present a proof-of-concept experiment at the DRACO~PW laser in a pitcher-catcher configuration, stably producing 6--\qty{7e7}{neutrons/shot} with energies above \qty{1}{MeV}, over more than 200 shots delivered at a shot-per-minute rate. Neutron time-of-flight measurements were performed using a single-crystal diamond detector, which is located only \qty{1.5}{m} away from the source and capable of resolving individual neutron-induced reactions. Observed reaction rates are consistent with Monte Carlo simulations inferred by real-time diagnostics of accompanying gamma, ion, and electron fluxes. With the recent advances in repetition rate, targetry, and ion acceleration efficiency, this work establishes LDNSs as a promising, scalable platform for future fast neutron-induced reaction studies, particularly for measurements involving short-lived isotopes or requiring high instantaneous neutron flux.
\end{abstract}

\maketitle
%
%
%


\begin{figure*}[htbp]
	\centering
    \includegraphics[width=1.\linewidth]{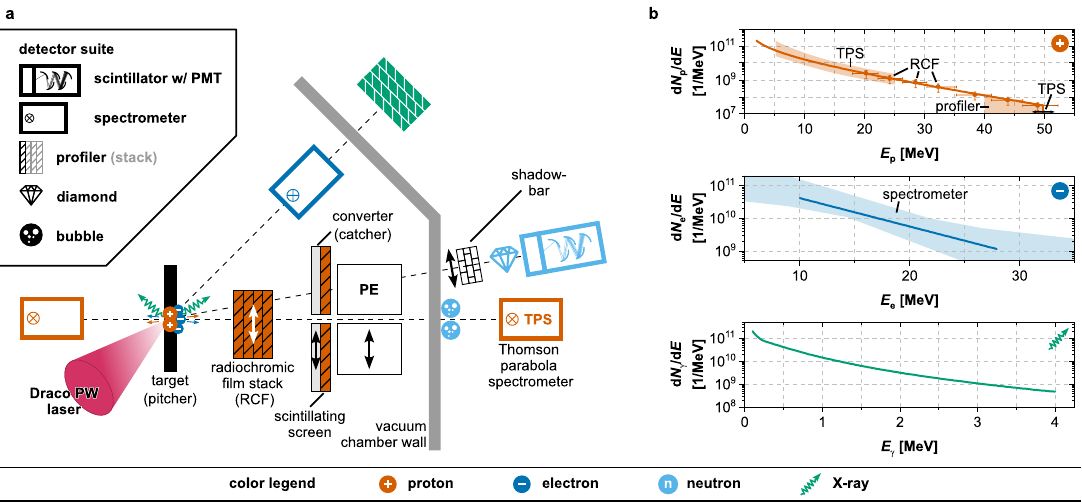} 
	\caption{(a) Experimental setup. The DRACO~PW laser is obliquely incident on the target. Proton/ion diagnostics (orange) are aligned along the target normal axis and comprise beam profilers (hatched boxes) and magnetic spectrometers. The electron spectrometer (dark blue) and bremsstrahlung calorimeter (green) are aligned along the laser forward axis. Diamond and neutron time-of-flight scintillators coupled to a photomultiplier tube (PMT) (light blue symbols) are aligned at $\sim\qty{30}{\degree}$ from the laser axis. Bubble detectors (light blue) and radiochromic film (RCF) stacks were placed for dedicated shots. Behind the target, the neutron converter, the PE-block, and the shadow-bar in front of the diamond detector could be placed or removed as desired. (b) Simulation input. Solid lines in the panels represent interpolations of the absolute measured spectral data for protons (top), electrons (middle), and bremsstrahlung (bottom) used as input for the MC simulations. The shaded areas refer to the data regions monitored on a single-shot basis with the corresponding detectors, as indicated by the symbols (see also Fig.~\ref{Fig_TPSanalysisANDstability}).}
	\label{Fig_ExperimentalSetUp}       
\end{figure*}

\section*{Introduction}

Neutron-induced reactions play a crucial role in various research fields
, from exploring the origins of elements in our universe~\cite{Kappeler2011} to understanding the core principles of nuclear physics~\cite{MATARRANZ201376}. These reactions are also relevant to dosimetry \cite{GOMEZROS2023106908}, medicine~\cite{ Banaee2021, Jin2022}, space science~\cite{Al-Khafaji2023}, material science~\cite{jeffries2021small}, condensed matter physics~\cite{mezei2022very}, and nuclear fission and fusion technology \cite{FORREST2011}, highlighting their broad scientific, societal, and industrial relevance.

As many research reactors are decommissioned, particle accelerators take the lead in experimental neutron science. To study nuclear structures, fission dynamics, material composition, and nucleosynthesis processes at a high resolution and across a broad neutron energy range, accelerator-based neutron sources commonly employ the neutron time-of-flight (ToF) method: pulsed neutron beams are generated and travel a well-defined flight path to a sample where they induce nuclear reactions. To measure, e.g., neutron-energy-dependent reaction cross-sections, individual reactions can be detected via promptly emitted radiation. This enables determination of the flight time---and thus kinetic energy---of the interacting neutron with the energy resolution primarily governed by the initial neutron pulse duration and the flight path length.

Nuclear reactions induced by fast neutrons ($>\qty{1}{MeV}$) are of particular importance for fission, fusion, and dosimetry, as reflected in the Nuclear Energy Agency’s High Priority Request List, where they account for over \qty{90}{\percent} of the requested nuclear data \cite{HPRL}. Additionally, more than \qty{25}{\percent} of requests concern radioactive actinide isotopes. The same applies to neutron capture in astrophysics \cite{Kappeler:2011}, in which the \textit{s}-process branching points of interest are all radioactive. Here, the pulse intensity is particularly advantageous for measurements on said radioactive samples as it improves the ratio between signal and radioactive background \cite{Colonna2018}. As a result, there is increasing interest in exploring alternative sources that can deliver intense, energetic neutron pulsees to address these experimental needs.

Even though the world's largest ion accelerators host proton spallation facilities that provide such neutron pulsees at $\sim\unit{Hz}$ repetition rates \cite{Guerrero:2013, Weiss2015, Reifarth:2005, Tang:2021, Kino:2011}, the nuclear physics community is looking to expand the range of neutron sources to meet current and future demands. In this context, advanced compact ion sources based on laser-plasma acceleration (LPA)---a rapidly advancing technology---have been discussed as drivers of fast neutron sources to complement and extend the existing infrastructure for nuclear physics research using the neutron time-of-flight method \cite{Pomerantz2014}. 

In LPA, an ultra-short (\unit{fs} to \unit{ps}) high-power (\unit{TW} to \unit{PW}) laser pulse interacts with a solid target, generating ultra-short ($\sim\unit{ps}$) ion pulsees in a plasma process accompanied by a strong prompt emission of electrons, X-rays, and electromagnetic noise. The acceleration process favors ions with a high charge-to-mass ratio, in particular protons. LPA proton pulsees contain up to \num{e12} protons at \unit{MeV} energies and exhibit a broad angular distribution and an exponentially decaying energy spectrum with a cut-off towards higher kinetic energies. It has recently been demonstrated that petawatt-class (\unit{PW}) lasers operating at $\unit{Hz}$-level repetition rates can reliably produce intense proton pulsees of several tens of \unit{MeV} \cite{Ziegler2021, Kroll2022}.

Owing to the characteristics of LPA beams, the corresponding laser-driven neutron sources (LDNSs) offer an ultra-high instantaneous flux (e.g., \num{e11} neutrons/shot \cite{Yogo2023}), which presents a dual advantage for nuclear reaction studies with neutron time-of-flight: First, for a given energy resolution, their short pulse duration allows for a short flight path length and therefore compact setup dimensions. Second, as discussed above, a high instantaneous number of neutrons is particularly advantageous for measurements on radioactive samples.

LDNSs have been experimentally realized and investigated, most commonly by directing LPA protons onto neutron converter targets in various configurations \cite{roth2013bright, Lelievre2023, Jiao2023} (see Ref. \cite{Mirfayzi:2025} for the most recent review). Proof-of-concept experiments have shown the use of laser-driven neutrons for radiographic imaging \cite{Zimmer2022, Yogo2023}, material analysis \cite{Mirani2023}, and as diagnostic tools in laser-plasma physics \cite{Alejo2017, Alejo2022,Yao2023}. 

However, although nuclear physics experiments with LDNS are recognized as a key scientific case at upcoming world-class laser facilities like ELI-NP \cite{ELI-NP:2016}, their feasibility has yet to be experimentally demonstrated. To show that a future research program on neutron-induced nuclear reactions with LDNSs is viable, three key capabilities must be established:

(1) It is necessary to obtain as complete information as possible about individual reactions and interaction events, which necessitates single-event detection in the harsh environment of an LPA, i.e., neither passive neutron detectors, such as bubble detectors \cite{Ing1997}, nor retrospective activation methods \cite{Mori:2021}, nor the typical current-mode operation of neutron scintillator detectors \cite{mirfayzi2015calibration}---as commonly employed in LDNS experiments---are suitable, since these detectors integrate over all interactions within their active volume.

(2) To achieve sufficient statistics, single-event signals need to be accumulated over many laser shots, requiring stable, high-repetition-rate source operation and shot-to-shot monitoring of experimental conditions.

(3) Accurate background characterization and subtraction, which is especially critical in LDNS environments. Unlike conventional accelerator-based sources, laser-plasma interactions generate an intense, prompt, and mixed background---including photons, electrons, ions, and scattered or secondary neutrons---whose properties are not yet fully understood or standardized. 

These capabilities are demonstrated simultaneously for the first time in this work (see Refs. \cite{Millan:2023, Scheuren:2024} for more details). The experiment discussed herein established the feasibility of studying fast neutron-induced nuclear reactions via neutron time-of-flight using a petawatt-class, ultrashort-pulse laser source operating in a stable multi-shot mode. 

A suite of dedicated particle diagnostics enabled shot-to-shot monitoring of neutron, electron, proton, ion, and photon production, demonstrating the required source stability. In parallel, Monte Carlo (MC) simulations were used to model the influence of background radiation components at the detector position, which were also measured. A high-performance diamond detector was placed at only \qty{1.5}{m} flight path length, maximizing neutron flux while maintaining sufficient neutron energy resolution.

Signals from individual fast neutron-induced reactions were successfully detected over hundreds of laser shots. The measured time-of-flight spectrum was found to be in agreement with MC predictions based solely on the LPA beam diagnostics, overcoming long-standing limitations related to detector suitability, source stability, and background complexity in this challenging environment.


\color{black}

\section{Design and implementation of a time-of-flight setup for an LDNS}

\begin{figure}[htbp!]
\centering
        \includegraphics[width=1.\linewidth]{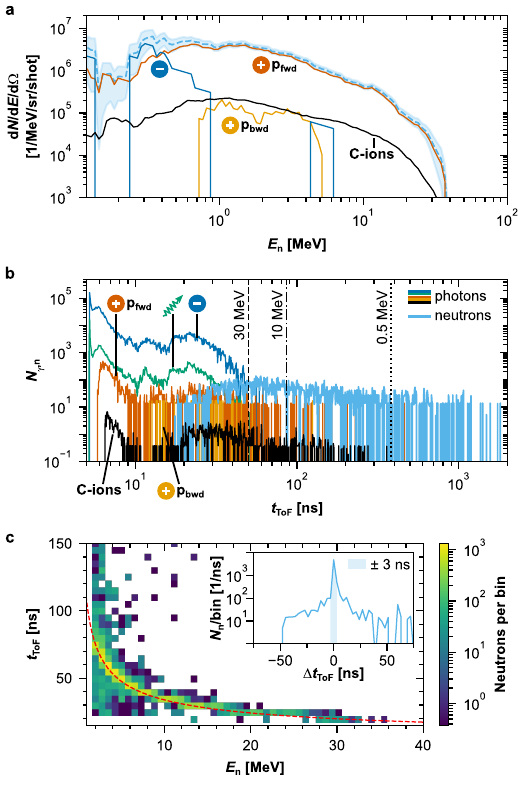} \\
    \caption{MC simulation results for particles arriving at the diamond detector placed \qty{1.5}{m} from the source. (a) The flux of neutrons as a function of neutron energy according to their origin. Forward-emitted protons ($p^+_{fwd}$) dominate neutron production in the high-energy region. The non-continuous spectrum for neutrons generated by electrons is an artifact of insufficient statistics. The dashed light blue line corresponds to the sum of the individual neutron production channels. The shaded area represents the statistical uncertainty of the simulation results for a given energy bin. (b) Time of arrival of prompt photon radiation fields (predominantly from bremsstrahlung from the source and bremsstrahlung produced by electrons elsewhere) vs. time of arrival of neutrons. A shorter flight path would imply an overlap between the arrival of photons and high-energy neutrons. The vertical dashed/dotted lines mark different neutron energies. (c) Correlation between the time-of-flight and energy of the neutrons. All events far from the red-dashed line, representing primary neutrons, correspond to neutrons scattered or produced elsewhere by secondary sources other than in the catcher target.  The inset displays a histogram of the difference between the neutrons' arrival time at the detector and the expected time corresponding to the direct path according to their energy. Only the neutrons with differences of less than \qty{3}{ns} (width of the initial ion pulse, shaded area. See inset in Fig.~\ref{Fig_DRACOvsOtherTOFfacilities}a) are considered primary neutrons produced in the catcher.}
    \label{Fig2}    
\end{figure}

Traditional accelerator-based neutron time-of-flight facilities maintain a distinct spatial separation between the neutron source and the detection apparatus. Thick walls equipped with collimators and neutron beam guides define the neutron flight path towards the measuring stations and suppress photon and neutron scattering backgrounds from the neutron production site. This is essential for ensuring accurate and reliable experimental results but, for now, it remains impractical with current LDNSs, as their multi-purpose laser interaction rooms are not yet optimized for neutron time-of-flight measurements. This implies close proximity of the neutron detection systems not only to the neutron production target, commonly referred to as \textit{catcher}, but also to the laser-plasma interaction point, or \textit{pitcher}. The latter emits a strong pulse of electromagnetic radiation (EMP) in the entire spectrum (\unit{THz}, optical, X-rays, and hard $\gamma$-rays) as well as particles (electrons, protons, heavy ions) across a large solid angle, some of which are energetic enough to produce neutrons and photons not only in the catcher but in other components of the experimental setup and the laser interaction chamber. To address these challenges, a streamlined setup was employed, focusing on three key tasks, as illustrated in Fig.~\ref{Fig_ExperimentalSetUp} to Fig.~\ref{Fig_TPSanalysisANDstability}: (1) Characterizing the laser-plasma interaction to feed MC simulations for (2) predicting neutron generation channels and detection rates in an optimized detector and shielding configuration; and (3) continuous shot-to-shot monitoring of particle beams generated in the LPA process to enable data accumulation, as required by high-precision nuclear physics experiments.

The experiment utilized the DRACO Petawatt (PW) laser at Helmholtz-Zentrum Dresden-Rossendorf. The laser system is capable of delivering pulses of up to \qty{18}{J} (on target) with a pulse duration of \qty{30}{fs} \cite{Schramm_2017}. These pulses were directed on the pitcher target at an oblique incidence angle of \qty{50}{\degree} using an f/2.3 parabolic mirror, achieving a peak intensity of $\sim\qty{5e21}{W/cm^2}$. To improve the laser pulse's temporal contrast, and thereby optimize acceleration performance and stability, a single plasma mirror setup was employed \cite{Ziegler2021}. This, however, restricts the experiment's repetition rate to about one shot every ten seconds and the daily shot count to approximately 200. The experiment employed plastic foils with a thickness of about \qty{250}{nm} as pitcher targets, with protons and carbon ions being the most abundantly accelerated species. These ions were emitted in normal directions from both the rear and front sides of the target through the well-established Target Normal Sheath Acceleration (TNSA) mechanism \cite{Snavely2000}.

The ion beam generated at the target's rear was characterized using a suite of detectors, each based on different detection principles, to accurately determine the respective contributions responsible for the neutron generation in the catcher (details in the Methods section). Central to this setup was a Thomson Parabola Spectrometer (TPS) aligned with the target's normal direction to analyze the particle spectra coming from the rear surface of the pitcher (forward direction, fwd) with high energy resolution. It featured a microchannel plate readout for cut-off energy detection and a Gd\textsubscript{2}O\textsubscript{2}S:Tb (GOS, on plastic support layer) scintillating screen for the absolute quantification of particle numbers \cite{Schilz:2024}. To complement the spectral analysis, another GOS scintillating screen was positioned $\sim\qty{10}{cm}$ behind the pitcher target to measure lateral beam profiles of accelerated protons above a discrete threshold energy of \qty{40}{MeV}. In tandem with the TPS, this profiler was used for the day-to-day optimization of the beam, adhering to the protocol established in Ref. \cite{Ziegler2021}. A magnetic spectrometer adapted from \cite{Jung2011} served to estimate the proton cut-off energy coming from the front side (laser incident side) of the pitcher (backward direction, bwd) by differentiating between protons and heavier ions with aluminum filters. For selected shots, stacks of radiochromic films (RCFs) were placed to map the energy-resolved spatial dose distribution delivered by the proton beam. Electrons and bremsstrahlung radiation were measured along the laser propagation direction. This was achieved by combining a magnetic electron spectrometer equipped with GOS screen readout inside and an imaging-plate-based bremsstrahlung calorimeter \cite{lasoGarcia2022,Prencipe2021} placed outside the vacuum chamber. The calorimeter featured a sequence of absorber plates made of different materials with increasing thicknesses, enabling the formation of a secondary particle shower recorded by the IPs as active layers.

Fig.~\ref{Fig_ExperimentalSetUp}b summarizes the particle beam spectra retrieved from the detector suite and used as input for the MC simulations. The proton energy spectrum directed towards the catcher is shown in the top panel. It exhibits the typical quasi-exponentially decaying spectral shape with well-defined cut-off energy of about \qty{50}{MeV}, surpassing by far the \qtylist{1.9;2.2;4.2}{MeV} thresholds of the \textsuperscript{7}Li, \textsuperscript{63}Cu, and \textsuperscript{65}Cu (p,n) reactions. With yields of about \qty{5e11}{protons/shot} for energies over \qty{2}{MeV}, the setup enables prolific fast neutron generation. Regarding carbon ions (not shown), cutoff energies of $E_\mathrm{max} \sim \qty{13}{MeV/u}$ were measured, and a total number of $\sim\num{3e9}$ ions ($> \qty{1}{MeV}$) was estimated for use in the simulations by scaling the energy deposition measured in the detector according to the stopping power difference between protons and carbon ions; a conservative estimate was adopted, as the detector was not absolutely calibrated for these species. The contribution of oxygen ions was not examined but deemed negligible for this analysis. The spectrum of particles emitted backward from the front face of the pitcher target is calculated from the RCF measurement from the rear side and the measured cutoff energy range in the magnetic spectrometer of about \qty{35}{MeV} (not shown), assuming again a TNSA scenario \cite{Poole2018}. During the laser-plasma interaction, electron and bremsstrahlung emission is strongest in the laser-forward direction, where respective energy spectra were measured (middle and bottom panels of Fig.~\ref{Fig_ExperimentalSetUp}b). To estimate their contribution to the neutron background, the simulations assumed isotropic emission characteristics and conservatively scaled the absolute numbers of electrons and photons, based on experimental measurements.\\

Neutrons were generated using two different catcher configurations: \qty{3}{mm} thick Cu and \qty{10}{mm} thick LiF. Inside the catcher material, fast neutrons are mainly produced via (p,n) reactions and emitted isotropically with \unit{MeV} energies, featuring an exponential behavior spectrally, inherited from the incident proton beam. 
In order to study neutron-induced nuclear reactions---particularly via time-of-flight---it is essential to detect individual interaction events and extract relevant observables such as timing and deposited energy. This requires detectors capable of operating in single-event mode with high temporal resolution. 

In previous LDNS experiments, detection systems have mostly operated in current or integral mode, or relied on passive detectors such as activation foils or bubble dosimeters. While these approaches are suitable for estimating total neutron yields or monitoring source stability, they do not provide the time-resolved and event-specific information required for neutron spectroscopy or reaction studies.

A detector suited to this task must combine fast response (to resolve signals separated by nanoseconds), radiation hardness (to withstand the prompt electromagnetic and particle emissions), and insensitivity to gamma and X-ray backgrounds (to avoid blinding or pile-up effects). Additionally, low intrinsic detection efficiency is beneficial to minimize pile-up in high-flux conditions.

To meet these requirements, a single-crystal diamond detector was selected and deployed in time-of-flight configuration (marked in light blue in Fig.~\ref{Fig_ExperimentalSetUp}a). These detectors are known for their sub-\unit{ns} response time, robustness under radiation, low sensitivity to photons, and proven suitability for neutron-induced reaction studies in harsh environments. Indeed, these detectors have previously been employed successfully in ToF nuclear physics experiments involving neutron-induced reactions, such as (n,$\alpha$) and fission \cite{Weiss2013, Weiss2016}. Their demonstrated performance in demanding radiation environments---such as the n\_TOF-NEAR station at CERN \cite{diakaki2022} and the planned neutron diagnostics at ITER \cite{diamond4fusion, WEISS_fus}---supports their suitability for use in LPA sources as well.  

For shot-to-shot neutron monitoring, a conventional EJ-232Q liquid scintillator \cite{mirfayzi2015calibration} coupled to a PMT was placed at a distance of \qty{3.6}{m} from the source and shielded from the prompt radiation by an up to \qty{15}{cm} thick assembly made of lead and polyethylene ---a common approach with LDNSs. Although quite limited for the accurate characterization of the neutron beam, the integral of the detector signal in current mode maintains proportionality to the overall neutron production per shot and thus may serve as a figure of merit to assess the shot-to-shot stability of the neutron source. Last, standard neutron dosimeter bubble detectors from Bubble Technology Industries \cite{BubblesTech} were used to determine the fast neutron yield.\\

\textbf{Secondary neutron sources}. Contrary to conventional accelerator-based sources, particles other than protons are also produced in the laser-pitcher interaction and are emitted in different directions. The impact of these is strongly dependent on the specifics of the experimental setup and has been neglected in all previous studies. In this work, their contribution was assessed using MC simulations. As illustrated in Fig.~\ref{Fig2}a, it turns out that a small but non-negligible fraction of the neutrons originates from reactions other than proton-induced interactions in the catcher. This secondary contribution is dominated by carbon ions for neutrons above a few \unit{MeV}, and electrons for neutrons below \qty{1}{MeV}. \\

\textbf{Photon background}. Since all charged particles are absorbed either by the catcher, the wall, or the shielding, only direct or secondary photons can reach the detector and induce additional signals to those from neutrons. Their contribution at the diamond detector position is displayed in Fig.~\ref{Fig2}b alongside the simulated neutron yield: the photon background, produced predominantly by electrons, is only significant at short flight times, i.e, in the region just before the arrival of the highest-energy neutrons (over \qty{30}{MeV}). \\
    
\textbf{Scattered (indirect) neutrons}. A crucial aspect in ToF experiments is the fraction of neutrons reaching the detector indirectly after being scattered or produced in secondary processes by the surrounding elements of the setup. Kinetic energies and arrival times of the neutrons reaching the diamond detector were simultaneously scored, and the corresponding heatmap plot is displayed in Fig.~\ref{Fig2}c, on top of which the dashed red line indicates the expected relationship for neutrons traveling without interactions from the catcher to the detector. All signals far from this kinematic region can be attributed to scattered or secondary neutrons, i.e. neutron background. The direct and background components can be distinguished better in the inset: all signals differing from the expected behavior by more than the \qty{3}{ns} duration of of the ion pulse (see Fig.~\ref{Fig_DRACOvsOtherTOFfacilities}a) can be attributed to background. According to the simulations, in our setup, the fraction of direct neutrons at the diamond detector is still dominant over the background at \qty{65}{\percent}. In comparison, the fraction of signal versus background amounts to \qty{48}{\percent} for the bubble detector (closer to the catcher but surrounded by structural material, and more sensitive to background due to its relatively flat efficiency down to low neutron energies), and only \qty{9}{\percent} for the scintillator (further from the catcher, surrounded by massive shielding and close to the hall walls).
 

These observations highlight the complexity and limitations of using scintillator setups and integrating detectors as bubble detectors for neutron yield characterization or comparative studies between different LDNS, when discriminating signal origins is essential.

The simulation results in Fig.~\ref{Fig2} illustrate the key design trade-offs to consider when optimizing a time-of-flight setup. A shorter flight path generally increases the neutron flux while reducing the neutron energy resolution. In this context, the neutron flux in the detector needs to be high enough to provide a sufficient count rate to allow for a measurement of the reaction yield given the mentioned limit in the number of laser shots per day, but avoiding both signal pile-up and superposition of the $\gamma$-flash and the neutron signals. In this work, at the chosen distance, the simulations predict a reasonable energy resolution ---on the order of a few percent, assuming a neutron production time of approximately \qty{3}{ns}--- with only minimal overlap between photon and neutron events. The expected count rate is a few signals per shot in the diamond detector, which is sufficient to keep pile-up under control while still providing adequate statistics within the $\sim$200 shots-per-day limit imposed by the use of the plasma mirror.

\section{Shot-to-shot monitoring}
Stability and reproducibility of the laser acceleration performance are essential for multi-shot nuclear physics experiments at LDNS, given that such sources exhibit fluctuations in energy and intensity. This has been a major concern for multi-shot applications of LPA, such as radiation therapy for cancer research, and can only be addressed through reliable shot-to-shot monitoring.

\begin{figure}[t!]
	\centering
	    \includegraphics[width=\linewidth]{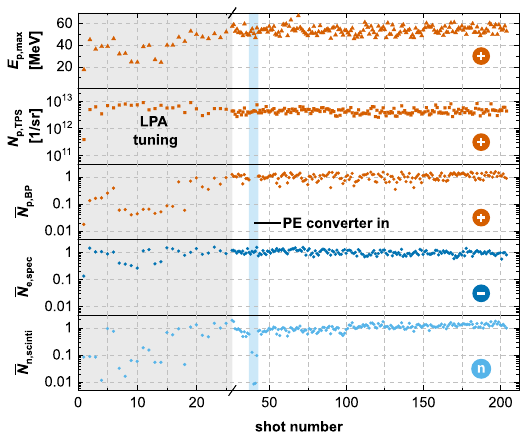}
	\caption{Shot-to-shot stability of protons (TPS cut-off energy $E_\mathrm{p,max}$, integrated proton yield from TPS $N_\mathrm{p,TPS}$ (\qtyrange{5}{25}{MeV}) and beam profiler $\overline{N}_\mathrm{p,BP}$ ($>\qty{40}{MeV}$)), electrons $\overline{N}_\mathrm{e,spec}$ as marked in Fig.~\ref{Fig_ExperimentalSetUp}b, and the resulting neutron production (integrated scintillator yield $\overline{N}_\mathrm{n,scinti}$) for 204 consecutive laser shots using the Cu-plate catcher configuration. Shaded areas mark the shot range where the laser-plasma accelerator was tuned (shots 1-26) and the PE-block was used as a beam dump (shots 38-42). The strong dip in $N_\mathrm{p,BP}$ during LPA tuning is because proton energies dropped below the energy threshold of the beam profiler. The unshaded region represents shots with optimized neutron performance (stability within \qty{26}{\percent} standard deviation). The quantities of the lower three rows are normalized to the mean values within the optimized range.}
	\label{Fig_TPSanalysisANDstability}       
\end{figure}

As illustrated in Fig.~\ref{Fig_TPSanalysisANDstability}, all employed diagnostics --- including proton cut-off energy, proton yield, and high-energy electron production --- indicate stable performance of the LPA source and subsequent neutron production under optimized conditions throughout a single shot day.

Quantitatively, the relative standard deviations (RSD) obtained across 204 consecutive shots were: \qty{7}{\percent} for the proton cut-off energy ($E_\mathrm{p,max} =54(4)$~MeV), \qty{24}{\percent} for the proton yield ($N_\mathrm{p,TPS} = 4(1) \times 10^{12}$ p/shot), \qty{26}{\percent} for the proton beam profiler signal, \qty{19}{\percent} for the high-energy electron signal, and \qty{26}{\percent} for the neutron signal measured by the scintillator detector. These values demonstrate that, despite the inherent shot-to-shot fluctuations in laser-plasma acceleration, the system maintained adequate stability to enable the accumulation of statistically significant single-event neutron data.

Two dedicated subsets of shots indicated in Fig.~\ref{Fig_TPSanalysisANDstability} further demonstrate that the scintillator-based neutron diagnostic can effectively assess LDNS stability: (i) inserting a PE beam dump that fully absorbs the ion beam (eliminating neutron production), which leaves only a residual background signal on the scintillator, and (ii) during the tuning phase of the LPA (for best focus and optimized spectral phase, following the protocol in \cite{Ziegler2021}), where the neutron signal closely follows the evolution of the primary particle beams.

It is important to note that there is no widely established benchmark for these RSD values in high-power laser plasma acceleration experiments, as traditionally such experiments have been conducted in a single-shot regime with extensive preparation for each individual shot \cite{roth2013bright}. Typically, only one or very few shots per day were performed. Therefore, the significance of our reported RSDs lies not necessarily in their absolute magnitude, but rather in the ability to maintain consistent and repeatable fluctuations across hundreds of consecutive shots. This stable, multi-shot operation enables effective shot-by-shot monitoring, allowing the identification and exclusion of non-optimal shots or experimental drifts, which is essential for accumulating statistically meaningful data in nuclear physics applications.

\section{Single event based neutron time-of-flight spectroscopy}
In this work, a fast neutron and particle detector with low efficiency and low gamma sensitivity --- a single-crystal diamond detector from \textit{CIVIDEC Instrumentation} \cite{CIVIDEC}--- was successfully operated at only \qty{1.5}{m} from the LDNS target. This short flight path, chosen to maximize the neutron flux, represents a challenging scenario due to the intense prompt electromagnetic and particle background generated by the laser-plasma interaction. Despite these conditions, the detector was able to register signals from individual interactions of fast neutrons with the carbon atoms in the detector material \cite{Weiss2016}, demonstrating the feasibility of single-event spectroscopy in the immediate vicinity of a petawatt-class neutron source.

\begin{figure*}[htbp]
    \centering
        \includegraphics[width=0.495\textwidth]{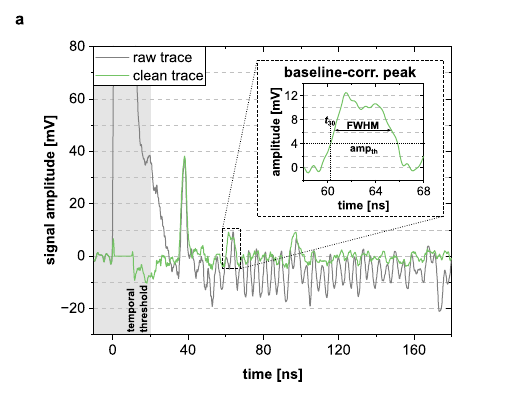}
        \includegraphics[width=0.495\textwidth]{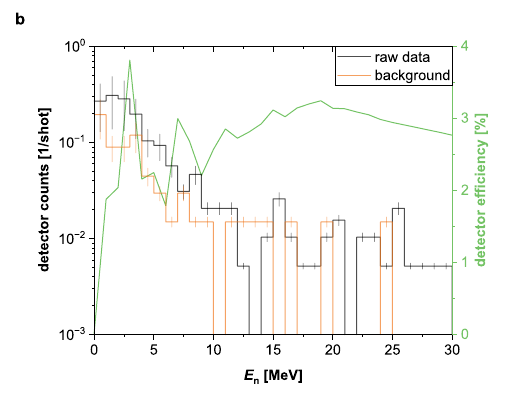}\\
    \centering
        \includegraphics[width=0.495\textwidth]{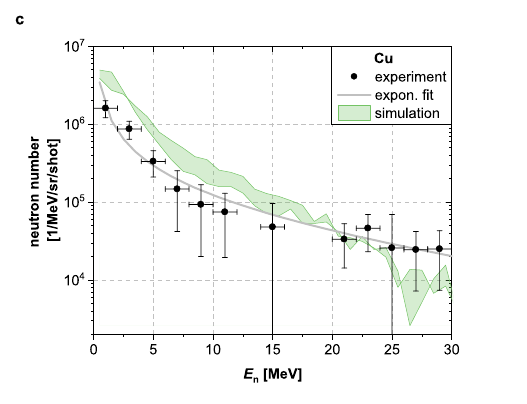}
        \includegraphics[width=0.495\textwidth]{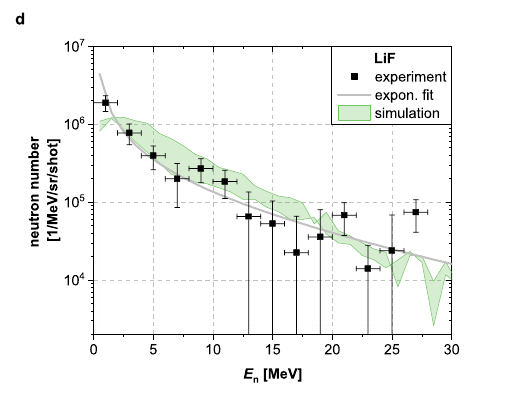}\\
    \caption{
    (a) Exemplary trace from the diamond detector before and after processing. The inset shows the notation of the signal characteristics for a selected single neutron signal. 
    (b) Neutron energy spectra for p-Cu pitcher-catcher configuration (black) and shadow-bar measurement (orange) showing the effect of the background. In green, the diamond detector efficiency is plotted considering the established \qty{400}{keV} deposited energy threshold. (c and d) Neutron yield obtained for Cu(p,n) (c) and Li(p,n) (d) after shadow-bar subtraction and taking into account the detector efficiency from (b). The exponential fit of the experimental data is shown as a solid gray line. Simulations are displayed as a band between Geant4 (lower limit) and PHITS (upper limit), estimating the uncertainty associated with the simulations.
    }
	\label{Fig_Diamond}       
\end{figure*}

A trace from the diamond detector corresponding to a single laser shot is shown in Fig.~\ref{Fig_Diamond}a, obtained after optimizing the acquisition conditions. The plot illustrates the trace before (raw trace, grey) and after applying quality improvements by subtracting the signal of the $\gamma$-flash, the corresponding baseline undershoot, and the EMP-induced noise (green trace, details in Methods). Still, a minimum time-of-flight threshold of \qty{20}{ns} remains necessary in the analysis, rejecting all signals with shorter arrival times because they overlap with the residual dominant $\gamma$-flash structure (indicated by the shaded gray area in Fig.~\ref{Fig_Diamond}a). This corresponds to a maximum detectable neutron energy of \qty{30}{MeV}, consistent with the expectations from the Monte Carlo simulations shown in Fig. \ref{Fig2}b. The remaining EMP-related fluctuations were still comparable to the amplitude of the neutron signals, necessitating the development of a dedicated pulse Shape Analysis (PSA) algorithm. This algorithm identifies and analyzes neutron-induced reaction signals in the diamond, considering as candidates positive signals with an amplitude larger than two times the noise standard deviation, equivalent to \qty{4}{mV} amplitude. These were analyzed and characterized by (see inset) amplitude, integral, width (FWHM), and timestamp (time corresponding to \qty{30}{\percent} of the amplitude in the rising slope), i.e., the time-of-flight, which is then converted into the corresponding kinetic neutron energy. Care was taken upon the occurrence of multiple-peak structures, which were statistically examined to discard noise peaks superimposed on actual signals that would then exceed the threshold.\\

To determine the neutron spectrum generated at DRACO~PW, signals from individual neutron interactions were measured using the diamond detector and accumulated for two catcher materials: LiF and Cu. Lithium was the natural choice, as it is commonly used in accelerator-based facilities operating in similar energy ranges \cite{Feinberg2009, Ledoux2021, MillanCallado2024}. Copper, on the other hand, was selected as an alternative configuration to support the validation of the simulation results.\\

The corresponding time-of-flight spectrum from each catcher was converted into neutron energy distributions after background subtraction and correction for the neutron detection efficiency. For background estimation in the detector due to the indirect neutrons discussed before, a shadow-bar (\qty{10}{cm} wide and \qty{40}{cm} long block of borated polyethylene) was inserted in the line of sight between the neutron source and the detector, acting as an efficient neutron shield \cite{shadowbars}. In Fig.~\ref{Fig_Diamond}b, raw and background detector signal histograms are compared for the Cu-catcher configuration. According to the shadow-bar method, neutron scattering accounts for \qtyrange{40}{50}{\percent} of the signal. This is consistent with expectations, considering the limited laboratory space and shielding constraints. The measured contribution matches the prediction of the simulation presented in Fig.~\ref{Fig2}c within a \qty{10}{\percent} (see the section on "Scattered (indirect) neutrons"). The neutron detection efficiency (green line in Fig.~\ref{Fig_Diamond}b) depends on the neutron energy and the signal amplitude threshold, in which the structures visible around \qty{5}{MeV} correspond to the resonances in the cross-sections for all possible \textsuperscript{12,13}C(n,*) reactions \cite{Weiss2016}. The efficiency curve is obtained based on Geant4 simulations \cite{agostinelli2003geant4} of the detector, considering mono-energetic neutrons in the range of \qtyrange{0}{50}{MeV} and counting the events with deposited energy above the detection threshold of \qty{400}{keV}. This value was selected as a compromise: sufficiently low to maximize the limited statistics, yet high enough to remain above the noise level.

After background subtraction and detection efficiency correction, the neutron energy distributions are then converted into neutron yield in the conventional units of \unit{n/sr/MeV/shot} by considering the number of laser shots delivered, the solid angle covered by the diamond, and the bin size of the histograms. The neutron yields for the Cu and LiF catchers are shown in Fig.~\ref{Fig_Diamond}c and d. In both cases, the neutron spectrum features a nearly exponential behavior, which agrees with the TNSA spectrum shape of the laser-driven proton spectrum.

For neutrons with energies above \qty{100}{keV}, yields of $\sim$\qty{1e8}{n/shot} were measured for both Cu and LiF catchers. Above \qty{1}{MeV}, the yields decrease to \qty{6e7}{n/shot} and \qty{7e7}{n/shot}, respectively. This nearly \qty{20}{\percent} difference is consistent with simulation predictions and can be attributed to the higher threshold of the Cu(p,n) reaction. The maximum neutron energy measured in both cases is $\sim$\qty{30}{MeV}, limited by the shortest time-of-flight accessible after the $\gamma$-flash and by reduced statistics at higher energies. Nonetheless, neutron production up to \qty{50}{MeV} is expected for the proton energy range of the primary beam (cf.\ Fig.~\ref{Fig_ExperimentalSetUp}b).

The neutron yields per shot reported herein are lower than, but still in good agreement with, bubble detector measurements, matching their order of magnitude and falling within a factor of 4 (see Ref.~\cite{Scheuren2024}). Furthermore, the diamond detector results for both catchers are consistent with the MC simulation predictions, as shown by the green overlays in Fig.~\ref{Fig_Diamond}c and d. This applies to the total neutron flux reaching the diamond detector, the spectral shape, and even the relative contribution of neutron scattering background, as measured using the shadow-bar technique. The level of agreement falls within \qtyrange{20}{50}{\percent}, which is noteworthy considering that each data set was obtained from fewer than 200 laser shots --—limiting the available statistics--— and that daily fluctuations in proton yield were already of comparable magnitude.

Overall, although no explicit nuclear reaction was investigated in this experiment, the registration and processing of single neutron-induced signals generated within the diamond detector already correspond to $^{12,13}$C(n,*) reactions produced within the detector material itself. In addition, the ability to identify individual signals and extract well-defined observables---such as time, amplitude, and FWHM (related to the incident energy, deposited energy, and signal shape, respectively)---opens the door to pulse-shape discrimination analysis and enables the exploration of correlations between these observables for deeper insight into the reaction processes of interest. These findings demonstrate the feasibility of studying neutron-induced nuclear reactions via time-of-flight using a PW laser-driven neutron source.

\section{DRACO LDNS in the landscape of ion accelerator-based neutron sources for nuclear reactions}

The main features of a neutron time-of-flight facility are neutron energy range, energy resolution, and neutron flux, either per pulse or average. In particular, LDNSs offer the potential of unprecedented neutron flux per shot and energy resolution thanks to the high-power and ultra-short duration of the laser pulse.

\begin{figure}[htbp]
    \centering
        \includegraphics[width=\linewidth]{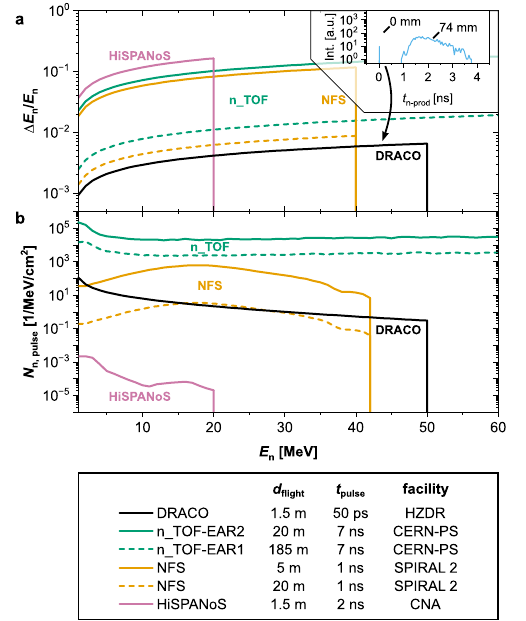}
    \caption{Performance of the DRACO LDNS and conventional ion accelerator neutron beam facilities. (a) The exceptionally short ion pulse generated by the laser (see inset) results in a better resolution at only 1.5 meters than other state-of-the-art facilities featuring flight paths as large as tens and even hundreds of meters. (b) The neutron yield per pulse varies enormously between facilities, featuring the DRACO LDNS orders of magnitude higher values than current compact accelerator-based neutron sources (CANS, i.e., HiSPANoS), but, as NFS, orders of magnitude lower than the spallation-driven ToF facilities.}
    
	\label{Fig_DRACOvsOtherTOFfacilities}       
\end{figure}

Regarding the excellent neutron energy resolution, it is related to the \unit{fs} duration of the laser pulse that produces ion pulses lasting only a few \unit{ps} \cite{Dromey:2016}. The most energetic of these ions, main responsible of the neutron production, spend only a few tens of \unit{ps} traversing the catcher target, producing a neutron burst of the same temporal scale, up to $\sim\qty{50}{ps}$ in the current experimental setup. In this work due to the \qty{74}{mm} separation between pitcher and catcher, required for parallel characterization of the primary sources, the neutron pulse duration was further broadened to a few \unit{ns}; however, this distance can be easily reduced to near contact to significantly shorten the neutron pulse duration.
As illustrated in Fig.~\ref{Fig_DRACOvsOtherTOFfacilities}a, a \qty{50}{ps} neutron pulse at DRACO~PW (best case scenario) yields an energy resolution better than \qty{1}{\percent} over the entire energy range using a flight path of \qty{1.5}{m}. This performance surpasses that of current state-of-the-art ion accelerator-driven ToF facilities, including compact (HiSPANoS@CNA, \qty{2}{ns} bursts) \cite{MillanCallado2024}, medium-scale (NFS@SPIRAL2, \qty{1}{ns} bursts) \cite{Ledoux2021}, and large-scale setups (n\_TOF@CERN, \qty{16}{ns}) \cite{Guerrero:2013, Weiss2015}, many of which require flight paths of tens to hundreds of meters. This means that the ToF setup associated with an LDNS for nuclear physics can remain extremely compact, in line with the advantages of laser-driven accelerators.\\

Regarding neutron source intensity, Fig.~\ref{Fig_DRACOvsOtherTOFfacilities}b shows the neutron yield per pulse for the same facilities as discussed above. Notably, there is an eight-order-of-magnitude difference between a compact source using a modest \qty{3}{MV} tandem accelerator (HiSPANoS@CNA) and the record-high flux achieved via \qty{20}{GeV} proton spallation at CERN n\_TOF. In this landscape, the intensity per pulse of the DRACO LDNS is comparable to that of the NFS facility at SPIRAL2 driven by a \qty{40}{MeV} deuteron beam, with neutron flight paths of \qtylist{5;20}{m}.\\

Achieving statistically significant results is essential for the success of nuclear physics experiments at LDNSs. To fully harness the potential of LDNSs, PW-class lasers operating at repetition rates of several \unit{Hz} are essential, considering that low repetition rates are only common in spallation sources, while lower energy accelerator-driven neutrons sources (for instance HiSPANoS or NFS) operate in the MHz regime. In this regard, results such as the recent demonstration of \qty{10}{Hz} operation at ELI-ALPS for comparable beam powers \cite{Nagymihaly:2023} are highly promising. Access to high repetition rates also demands advanced pitcher target solutions that enable a significantly larger, or even unlimited, number of laser shots. Encouraging progress is being made with concepts such as automated target holders \cite{Peñas_Bembibre_2024}, tape targets \cite{Xu:2023}, self-replenishing cryogenic jets \cite{Rehwald:2023}, and liquid leaf jets \cite{Treffert:2022, Streeter:2025}. Additionally, laser ion acceleration efficiency is advancing rapidly: for example, recent experiments at the DRACO facility achieved higher proton fluxes and proton energies exceeding \qty{100}{MeV}, notably without using the shot-rate-limiting plasma mirror \cite{Ziegler:2024}.\\

These developments, together with our proof-of-concept study on single-event-based neutron spectroscopy, support the conclusion that LDNSs---fueled by rapid progress in laser-plasma acceleration---are emerging as viable and complementary tools for neutron-induced nuclear reaction studies, particularly in scenarios where short pulse durations and high instantaneous fluxes provide critical advantages.\\

Having demonstrated the feasibility of measuring neutron-induced reactions, the accuracy at reach is still to be studied. Thus the next necessary step in this endeavor consists in measuring a nuclear reaction with well-known cross section. Assuming a 10-fold neutron production \cite{Ziegler:2024} and 10~Hz (as at ELI), cross sections in the order of 1 barn could be measured within a 3\% statistical uncertainty (1500 counts would be accumulated for each sample in \qty{1}{MeV} bins around \qty{10}{MeV}) in only one week of beam time employing a 4-fold 10x10~mm$^{2}$ mosaic diamond detector loaded with different samples. This should be enough to asses the accuracy at reach for nuclear reaction studies with the LDNS approach discussed herein. An interesting choice of  samples and reactions could be fission on $^{235}$U and $^{239}$Pu in the spirit of the very recent experiment of the NIFFTE Collaboration \cite{Dongwi:2025} at LANL
for measuring the $^{239}$Pu/$^{235}$U fission cross section ratio with an unprecedented accuracy of just 1\%.

\newpage
\section{Methods}
\textbf{Laser-driven neutron source:} 
Experiments were performed at the DRACO~PW laser at the Helmholtz-Zentrum Dresden-Rossendorf. DRACO~PW is a double-CPA Ti:Sa laser system, providing \qty{30}{fs} (FWHM) laser pulses with a central wavelength of \qty{810}{nm} (\qty{50}{nm} bandwidth FWHM). The pulses carried an energy of $\sim\qty{18}{J}$ and were focused onto thin ($\sim\qty{250}{nm}$) Formvar plastic foils (C\textsubscript{5}H\textsubscript{8}O\textsubscript{2}, $n_e = 230 n_c$ when fully ionized) by an f/2.3 off-axis parabolic mirror to a FWHM spot size of \qty{2.6}{\micro\meter} yielding peak intensities of \qty{5.4E21}{W/cm^2}. Plasma mirror (PM) cleaning was applied to improve the temporal contrast by almost four orders of magnitude, essentially removing all remaining prepulses and limiting the ionization and plasma dynamics to the last \unit{ps} before the peak of the laser \cite{Ziegler2021, Dover:2023, Bernert:2022, Bernert:2023}. PM operation limited the shot rate to about one shot every 20 seconds and the total number of shots to about 200 shots per day. Neutrons were generated using two different catcher options (\qty{10}{mm} thick LiF and \qty{3}{mm} thick Cu).\\

\textbf{Particle diagnostics:} 
The spatial proton intensity distribution was measured using calibrated radiochromic ﬁlm (RCF) stacks in several reference shots. A Thomson Parabola spectrometer (TPS) placed at \qty{45}{\degree} was used to measure ion spectra for every laser shot. A small fraction of the ion beam emitted from the target was delivered passing the \qty{4}{mm} diameter hole in the catcher and a pair of small pinholes towards the TPS. This resulted in a set of two parabolic traces, one recorded with a microchannel plate in imaging configuration to assess the cut-off energy with high spatial resolution and high dynamic range, and the other used a calibrated Gd\textsubscript{2}O\textsubscript{2}S:Tb (GOS) scintillating screen \cite{Schilz:2024} for absolute proton number determination in the proton energy range of \qtyrange{5}{25}{MeV}. The TPS and RCF data, with remarkable agreement, indicate that a total of \qtyrange{5e12}{6e12}{protons/sr/shot} are produced by each laser pulse. Considering the beam divergence of around $\sim$20${^\circ}$, this corresponds to a proton beam intensity of \qty{5.6e11}{particles/shot} (spectrally integrated for $>\qty{2}{MeV}$). The predominant heavy ion species was C$^{6+}$. In the text, these are labeled as carbon ions for simplicity. On-shot spatially resolved particle detection was conducted with a proton beam profiler. Another calibrated GOS scintillator screen was attached to the rear of the catcher to measure the beam profile of protons with an energy larger than \qty{40}{MeV}. A bandpass-filtered (\qty{540+-2}{nm}) CMOS camera captured the emitted luminescence light.\\

Electron spectra were recorded shot-to-shot using a magnetic spectrometer positioned in the laser-forward direction, \qty{640}{mm} from the source. Electrons entered through a \qty{10}{mm} diameter aperture and were magnetically dispersed and deflected onto a GOS scintillating screen imaged by a CMOS camera. Spectral information was retrieved using a custom code incorporating a General Particle Tracer simulation of the spectrometer response. The system was optimized for electrons with kinetic energies of \qtyrange{1}{60}{MeV} and allowed evaluation of electron temperatures in the range of \qtyrange{0.5}{40}{MeV}.\\

All neutron detectors were placed outside the vacuum chamber. Integral neutron production was measured using PND-type Bubble detectors and Bubble Reader from Bubble Technology Industries, positioned at various locations around the target area. To achieve reasonable bubble counts ($\sim100$, to reduce statistical and read-out errors), approximately 10 laser shots were accumulated. The bubble count was then converted into a neutron yield according to the protocol in \cite{Scheuren2024}, which considered the detector response from the manufacturer and applied temperature and energy corrections based on the MC-simulated neutron energy spectrum at the bubble detectors' positions.\\

Traces of the diamond detector (4x4~mm$^2$ surface and \qty{500}{\micro\metre} thickness), mounted at \qty{1.5}{m} from the target, were amplified using a \qty{2}{GHz} C2-HV Broadband Amplifier by CIVIDEC Instrumentation and recorded with a Tektronix MSO64B oscilloscope (\qty{6}{GHz} bandwidth, \qty{12}{bits}, \qty{10}{GS/s}). The main challenge was the prompt detector saturation due to the X-ray photon flash ($\gamma$-flash in the following) and the pile-up resulting from harsh radiation and high instantaneous flux, which prevented the measurement of individual fast neutron signals at low time-of-flight. These factors, the high signal originating from the $\gamma$-flash and the high interaction rate at the detector caused by the high instantaneous flux, generate the signal undershoot due to the coupling of the amplifier. To minimize these issues, \qty{50}{mm} of lead filter was added in front of the detector. 

Following the X-ray photon flash, the trace exhibited high-frequency noise for approximately \qty{250}{ns} due to the electromagnetic pulse (EMP) from the laser-plasma interaction. Initially, the EMP noise reached amplitudes up to \qty{200}{mV}, but it was reduced to few mV by using EM shielding around detector and electronics and by using passive low pass filters in the bias voltage supply cable. The detector was also electrically isolated to decouple its ground from the common ground of the whole accelerator installation. For this purpose, the detector was connected to an independent power supply, and plastic foils were used to prevent metallic contact with other surrounding materials. The oscilloscope was placed outside the laser bunker to prevent ground loops and extra EMP coupling in the acquisition system itself. The saturation due to the $\gamma$-flash and the EMP ringing were relatively consistent between shots of a single day, thus, the corresponding exponentially decaying tail and the constant oscillating component were averaged across all recorded traces and subtracted from each trace, resulting in a significantly cleaner baseline that allowed for better signal identification. Signal undershoot and associated baseline shifts were corrected by locally averaging the values immediately before and after the signal.\\

The detector was energy-calibrated to establish the detection threshold applied in the signal analysis. According to the manufacturer, the energy deposited in the detector for a signal depends directly on the area of the signal and the carbon ionization energy (\qty{13}{eV}), and inversely to the detector impedance (\qty{50}{\ohm}), the gain of the amplifier (150), and the electron charge. To avoid the integration of remaining noise in the signals, the area is calculated assuming the step-function shape of the signals \cite{Weiss2016}, i.e., multiplying the signal's width (FWHM) by its amplitude, being these variables are less affected by the noise. To assess the effect of the uncertainties in the signal area and possible gain variation, \qty{+-30}{\%} fluctuations in the calibration were tolerated, which are within the statistical uncertainties. \\

\textbf{Monte Carlo (MC) simulations:}
To evaluate the influence of the different secondary particles and to inform the design of the experimental setup, simulations were performed with the MC code PHITS (version 3.31) \cite{iwamoto2022benchmark, sato2023recent}, applying input parameters as presented in Fig.~\ref{Fig_ExperimentalSetUp}. The geometry was adapted to the real laboratory infrastructure consisting of concrete walls and floor, vacuum vessel (stainless steel and aluminum assembly) with inner components (such as glass parabola, breadboard, and magnet yokes), and detector system (bubbles, scintillators, and diamond), including applied lead shielding. Both catcher options, LiF and Cu, used in the experiment were evaluated. Within the simulation framework, the following settings were applied: (1) JENDL-4.0 \cite{shibata2011jendl, matsuda2019ace} nuclear data library was used for neutron transport and to model proton-induced nuclear reactions, e.g., (p,n) reactions. (2) TENDL-2019 \cite{koning2019tendl} was used for photo-nuclear reactions and (3) EGS5 (Electron-Gamma Shower, version 5) code \cite{hirayama2005egs5} for photon and electron transport. (4) The Kurotama \cite{iida2007formula} and Intra-Nuclear Cascade of Li\`{e}ge (INCL) \cite{boudard2013new} models were implemented as part of the simulation settings.
Simulations of the neutron efficiency of the diamond detector were performed for a simplified geometry with the GEANT4 toolkit \cite{agostinelli2003geant4, allison2006geant4, allison2016recent} using the QGSP\_BIC\_AllHP physics list, the JEFF-3.3 nuclear data library \cite{plompen2020joint} and the NRESP model \cite{garcia2017new} for neutron transport and proton-induced reactions. \\

\clearpage
\textbf{\Large Acknowledgements}.\\
We acknowledge the outstanding technical and experimental support of the Draco laser group.

The authors acknowledge financial support by: \\
- The Spanish Ministry of Economy and Competitiveness (RYC-2014-15271, FPA2016-77689-C2-1-R, RTI2018-098117-B-C21, PID2023-152894OA-I00 and RYC2021-032654I projects);\\
- MAMC  has received funding from the EC Euratom research and training programme 2014-2018 under grant agreement No 847594 (ARIEL) and the EC PEOPLE Marie-Curie Action programme 2007-2013 under grant agreement No 334315 (NeutAndalus); \\
- Xunta de Galicia (grant ED431F2023/21, CIGUS Network of Research Centres);\\
- “La Caixa” Foundation (ID 100010434), fellowship code LCF/BQ/PI20/11760027;\\
- María de Maeztu grant CEX2023-001318-M funded by MICIU/AEI/10.13039/501100011033S; \\
- The project was supported by Hi-Acts, an innovation platform under the grant of the Helmholtz Association HGF\\
- SSch acknowledges support from Trumpf SE + Co. KG;\\
- CR acknowledges support from the LOEWE excellence programme of the state of Hesse;\\

\textbf{\Large Author contributions}.\\
- MAMC, CG, TRG, BF, SSch, JK, CR, AA, JB, RB, AJ, FK, IP, SU, KZ, CW, JMN, MRe, TZ and EG prepared and/or conducted the experiments.\\ 
- MAMC and SSch performed the simulations. \\
- MAMC, SSch, CG, JK, FK, IP, KZ, JMN, RS, IP, MRe analyzed the data.\\
- CG, MAMC, SSch, FK, KZ, AJ wrote the manuscript. \\
- CG, FK, KZ, AJ, TEC and US supervised the project. \\

All the authors reviewed the manuscript and contributed to discussions. \\

\bibliographystyle{unsrt}
\bibliography{References}
\end{document}